# The role of ethical consumption in promoting democratic sustainability: revisiting neoclassical economics through Kantian ethics

*Pascal Stiefenhofer*

*Newcastle University, UK*

*pascal.stiefenhofer@newcastle.ac.uk*

**Abstract:** This paper examines the intersection of philosophical ethics and economic theory through the lens of ethical consumption, proposing its transformative potential to steer democratic governance toward sustainability. As awareness of environmental and social imperatives deepens, the demand for ethically produced and environmentally responsible products disrupts traditional economic paradigms focused on utility maximization and market efficiency. This shift, driven by increased transparency, equity concerns, and a commitment to values-based consumption, prompts a critical reassessment of foundational economic assumptions.

Conventional neoclassical models, with their emphasis on rational agents and market equilibrium, often neglect crucial elements of sustainability, including ecological integrity, social justice, and intergenerational ethics. In contrast, a democracy rooted in sustainability prioritizes ethical imperatives, long-term accountability, and equitable policies to foster a just future. This paper contrasts the neoclassical, market-centred approach with the growing influence of ethical consumers who prioritize sustainability and moral responsibility over sheer utility, arguing that embedding these ethical dimensions into economic frameworks is essential for creating a consumption paradigm grounded in inclusivity and ethical responsibility.

Drawing from White's Kantian-economic model of decision making (2011) and Inglehart's theory of intergenerational value transformation (1990), the paper illuminates how evolving societal values challenge traditional economic assumptions. Through a vector bundle approach with projection mapping, the paper proposes a model that integrates Kantian moral imperatives within economic structures, accommodating both fixed moral duties and flexible ethical preferences to capture value shifts over time. This approach envisions an economy aligned with post-materialist values, fostering ethical consumption and supporting sustainable societal well-being through a responsive, value-oriented democratic order.

## Introduction

In recent years, ethical consumption has gained considerable momentum, exemplified by practices such as *buycotting*, where consumers actively choose ethically produced goods, and *boycotting*, where they avoid products deemed unethical (Szmigin et al., 2009; Kam & Deichert, 2020). Rising environmental awareness has notably reshaped consumer preferences, driving demand for sustainable and ethically sourced products (Tan et al., 2019). Ethical consumers are *informed consumers* who consciously use ethical labels to guide their purchasing decisions, seeking products that align with values such as sustainability, social justice, and fair labor practices (Nicholls & Opal, 2005; Freestone & McGoldrick, 2007; Papaoikonomou et al., 2011; Stiefenhofer, 2019, 2021a). According to Andorfer and Liebe (2012), labels like Fair Trade not only signal that a product meets certain ethical standards but also help build *trust* between consumers and producers, particularly in industries known for social or environmental concerns. The Nielsen Global Responsibility Report (2016) highlights an accelerating shift towards ethical consumption, revealing an increasing consumer *willingness to pay premiums* for environmentally sustainable products (Andorfer & Liebe, 2012; Tan et al., 2019).

Empirical research highlights a substantial rise in ethical consumption as a defining trend in contemporary consumer behavior, marking a fundamental shift in market dynamics. Unlike traditional consumers who prioritize price and quantity to maximize utility, ethical consumers are motivated by a broader set of considerations, including sustainability, human rights, and social justice (Freestone & McGoldrick, 2007; Cowe & Williams, 2000). Increasingly, factors such as environmental impact, labor welfare, and animal rights play a critical role in these purchasing decisions (Naderi & Van Steenburg, 2018). Naderi and Van Steenburg (2018) suggest that this trend is deeply rooted in *moral values passed down through generations*, emphasizing the ethical importance of supporting products that align with environmental sustainability.

Inglehart provides further empirical evidence of this *intergenerational value shift*, offering a theoretical framework that situates ethical consumption within a larger societal transformation. This growing trend reflects more than just individual preferences; it is part of a broad cultural evolution that Ronald Inglehart's theory of intergenerational value change helps explain. According to Inglehart, as societies become more affluent and secure, their values shift from prioritizing material concerns, such as economic stability, toward post-materialist values, including self-expression, autonomy, and environmental responsibility (Inglehart, 1971, 1977). This framework, which draws on Maslow's hierarchy of needs (1970), posits that in affluent societies, where basic material needs are already met, individuals increasingly prioritize sustainability and ethical consumption as they pursue higher-order values. Consequently, these ethical considerations become central to consumer choices, further embedding them into the societal fabric of affluent, secure nations.

Prior insights from Thorstein Veblen, particularly his work in *The Theory of the Leisure Class* (1899), both predate and complement Inglehart's theory of intergenerational value change. Veblen's concept of *conspicuous consumption* demonstrates how, in affluent societies, individuals use consumption to signal *social status* rather than simply meet basic needs. This perspective aligns with Inglehart's later analysis, which highlights the shift toward *post-materialist values* in wealthier societies, where consumption increasingly reflects ethical and environmental concerns rather than survival or economic necessity (1990, 2008). In contemporary society, ethical products—such as those marketed as sustainable—now function as status symbols, reflecting a trend that Stiefenhofer (2021b) and Stiefenhofer and Zhang (2022) explores further in a theory of *conspicuous ethics*. This paper aims to integrate the intersecting theories of status, intergenerational value change, and ethical consumption to bridge the gap between traditional neoclassical economic models and economic decision models that incorporate ethical considerations. In doing so, it highlights how evolving societal values are reshaping consumption patterns—an aspect that has often been overlooked.

A recent approach to understanding the complexities of ethical consumption behavior is articulated by White (2011), who links Kant's concept of *duty* to economic decision-making, establishing a philosophical framework that deepens this research's grasp of the shift toward ethical consumption and its broader implications for democratic sustainability. Distinct from Veblen's and Stiefenhofer's perspectives, which interpret consumption as a vehicle for social signaling and status, Kant's ethical framework emphasizes *actions grounded in universal moral principles*. The Kantian *categorical imperative* directs individuals to act according to maxims that could be universally applied, suggesting that ethical consumption choices—such as prioritizing fair-trade or sustainable products—are motivated by a sense of moral duty rather than by self-interest or social recognition (Kant, 1785). This framework asserts that ethical consumption is guided not by utility maximization but by an obligation to contribute to the common good. This perspective stands in stark contrast to neoclassical economic models, which traditionally interpret consumer behavior as driven primarily by the pursuit of individual utility. White's approach thus provides a unique lens, positioning ethical consumption as a duty-based response to global challenges. This framework aligns consumer actions with ethical imperatives, promoting a model of economic behavior that supports sustainability and democratic resilience in the face of evolving social and environmental needs.

White (2011), in line with Sen (1977) and Hausman and McPherson (1993), argues that ethical considerations—such as justice, fairness, and moral obligations—should form the very foundation of economic analysis. White's model introduces three *ranking axioms*, creating a structured framework that prioritizes moral imperatives above individual preferences. This formalization underscores the centrality of ethical duties in economic decision-making, a perspective echoed by Sen (1985) and Basu (2011), who assert that economic rationality transcends the restrictive scope of utility maximization. They contend that true rationality is deeply intertwined with moral commitments and ethical responsibilities, suggesting that the economic agent is as much a moral actor as a utility maximizer.

While Friedman (1953) advocates for viewing economics as an objective, descriptive science free from normative judgments, scholars like Marglin (2008) and DeMartino (2011) challenge this perspective. They assert that economic models must integrate ethical considerations to more accurately reflect the complexities of real-world decision-making. This aligns with White's Kantian framework, which integrates duty and moral obligation into economic decision-making, suggesting that ethical consumption reflects not just individual preferences but a deeper commitment to sustainability, fairness, and social justice—values that are increasingly central in post-materialist societies.

White's model gains added significance in the context of corporate social responsibility, where businesses must balance profit with ethical obligations to stakeholders. Hsieh (2004) extends Kantian ethics to corporate decision-making, asserting that companies hold both perfect and imperfect duties to employees, consumers, and the environment. This perspective underscores the importance of integrating duty and moral responsibility into economic behavior, especially when addressing moral dilemmas in business.

The shift toward ethical consumption signifies not merely a change in consumer preferences but a more profound, value-driven transformation in societal priorities. As post-materialist concerns like sustainability and social responsibility become central to consumer behavior, economic models—especially those grounded in Kantian ethics and the theories of Veblen (1899) and Inglehart (1990)—must adapt to reflect these ethical dimensions. White's Kantian-model of choice underscores that ethical consumption is motivated by a sense of duty rather than personal utility, contributing to both individual and collective well-being. As Stiefenhofer (2019) argues, this evolution in consumption patterns also marks a broader transformation in how economies and societies value ethics. It highlights the role of consumer sovereignty, where *price shifts* driven by *price-dependent ethical preferences* empower consumers to influence market dynamics, a concept still underexplored within the Kantian framework. This *integration of ethics into pricing mechanisms* strengthens the relationship between consumer choice and democratic sustainability.

This paper aims to establish new theoretical connections and distinctions between the neoclassical economic model and White's Kantian model of choice. By examining societal value shifts—such as the increasing emphasis on sustainability and ethical consumption—the paper investigates how these trends contribute to a more sustainable and democratic framework within consumption patterns. It explores how emerging post-materialist values, rooted in ethical imperatives, challenge the foundational assumptions of traditional economic models, especially those focused solely on utility maximization. Through this analysis, the paper bridges the gap between rational choice theory and moral duty, illustrating how ethical consumption reshapes consumer behavior by blending individual preferences with collective moral responsibilities.

In critically analyzing the axioms in White's Kantian-economic model of decision making, this paper delves into both their mathematical structures and philosophical foundations. By rigorously examining these core axioms, it seeks to advance a formalization of White's model that integrates intergenerational value shifts to address contemporary societal concerns, including sustainability and ethical consumption. This approach provides key insights into how White's model can be refined and expanded to better reflect evolving ethical priorities.

Furthermore, by comparing White's model with the neoclassical economic model, the paper underscores important distinctions, revealing how each framework addresses the interplay of rationality, moral duty, and consumer behavior in the context of changing societal values.

## The rise of the ethical consumer

Empirical research elucidates the ascendancy of ethical consumption as a transformative force within market dynamics, suggesting a substantial reconfiguration of consumer motivations and economic structures. Unlike the archetypal consumer, whose calculus is traditionally governed by considerations of price and utility maximization, the ethical consumer embodies an expanded ethical horizon, attuned to values of sustainability, human rights, and social justice (Freestone & McGoldrick, 2007; Cowe & Williams, 2000). For this emergent category of consumers, factors such as ecological impact, labor equity, and animal welfare are integral to the deliberative process, displacing mere cost-efficiency as the primary criterion for choice (Naderi & Van Steenburg, 2018).

The literature indicates that ethical consumers exhibit a pronounced willingness to incur additional costs or forego convenience, thereby actualizing their commitment to these ethical imperatives (Andorfer & Liebe, 2012; Tan et al., 2019). This evolving praxis reflects a confluence of consumer agency with moral resolve, wherein ethical commitments are not peripheral but foundational, actively molding consumption patterns. In this respect, the ethical consumer emerges as an agent of normative disruption, compelling a reassessment of consumption as a domain of moral agency and social responsibility.

Environmental concerns have crystallized as pivotal motivators in the rise of ethical consumption. The Nielsen's report (2016) revealed that 73% of global respondents would accept a premium for sustainably sourced products, underscoring a profound shift in the valuation of goods beyond traditional economic metrics. This trend is further substantiated by the UK's Ethical Markets Report (2024), which documents a remarkable escalation in ethical consumer spending, leaping from £17 billion in 1999 to over £141 billion in 2023. Such growth finds its most fervent advocates among Millennials and Gen Z, cohorts distinguished by a pronounced ethical sensitivity; research illustrates their propensity to actively boycott brands that fail to uphold eco-conscious practices. Additionally, Wells et al. (2019) identifies a substantial proportion of consumers inclined to pivot between brands on the basis of ethical considerations alone, signaling a reorientation of market allegiances grounded in moral valuation. Thus, environmental ethics are not merely adjunct to consumer choice but are increasingly foundational, reshaping market behavior and embedding ecological imperatives at the heart of economic exchange.

The growing prominence of ethical consumption has sparked significant academic interest, leading to a rapidly expanding body of research addressing themes such as fair trade practices and sustainable purchasing behaviours (Andorfer & Liebe, 2012, 2015; Bray et al., 2011; Papaoikonomou et al., 2011; Carrington et al., 2010). Building on this discourse, Gandjour (2024) explores the limitations of models of mental state attribution, particularly in the context of moral value judgments, offering a nuanced perspective to the field.

This burgeoning scholarship reflects the significance of ethical consumption not only as a transformative market phenomenon but also as a profound area of philosophical inquiry, illustrating how shifting societal values are reshaping the economic landscape and redefining consumer expectations in fundamental ways.

## Shifts in values and consumption behaviors

Ethical consumption is inextricably bound to personal values, which are themselves shaped by generational and sociocultural currents. Naderi and Van Steenburg (2018) argue that commitments to sustainability and fairness are not merely individual predilections but are often inherited across generations, thereby exerting a profound influence on contemporary consumer behavior. This intergenerational transmission aligns with the broader societal shifts delineated by Inglehart (1971, 1977), which posits that economic conditions indelibly shape evolving value systems. While older generations, molded by experiences of economic scarcity, adhered to "materialist" values emphasizing security and stability, younger cohorts increasingly gravitate towards "post-materialist" values—principles of autonomy, self-expression, and environmental stewardship. This generational contrast elucidates the interplay between economic context and ethical orientations, highlighting how evolving economic circumstances reframe ethical imperatives such as sustainability (Inglehart, 1990).

Inglehart's theory of post-materialism contends that as societies achieve greater economic affluence, collective values undergo a paradigmatic shift from material preoccupations—such as security and stability—towards ideals that prioritize self-expression, environmental stewardship, and an enhanced quality of life.

Post-materialist orientations, therefore, elevate concerns for human rights, ethical consumption, and ecological preservation above the pursuit of economic growth and material accumulation (Dietz et al., 2005; Dunlap & York, 2008; Pepper et al., 2009; Micheletti & Stolle, 2012; Norris & Inglehart, 2019). This reconfiguration of values reflects a profound transformation wherein ethical and existential dimensions increasingly frame social priorities, signifying a departure from the utility-maximizing imperatives of earlier economic paradigms.

Stiefenhofer's notion of "conspicuous ethics" (2019, 2021a, 2021b) provides a compelling theoretical extension to the post-materialist shift, positing that the ethical premium attached to certain goods functions as a marker of distinction, allowing individuals to accrue social status through the accumulation of moral capital. This conceptual framework is an evolution of Veblen's theory of conspicuous consumption, wherein luxury goods serve as symbols of wealth and prestige. Stiefenhofer & Zhang (2022) adapt this paradigm, suggesting that in post-materialist societies, ethically produced goods come to signify moral virtue, reflecting a form of social currency grounded in ethical rather than economic capital. This interpretation resonates with Inglehart's insights, which acknowledge that in post-materialist contexts, values such as environmentalism may themselves become instruments of social distinction, as individuals signal their ethical commitments through consumption that transcends mere material necessity.

## Three pillars of ethical consumption: choosing, rejecting, and advocating for a just market

Ethical consumption can be delineated into three primary modalities: positive, negative, and discursive. *Positive consumption* entails the deliberate selection of products aligned with ecological and ethical values—such as organic, fair trade, or sustainably sourced goods (Lang & Gabriel, 2005; Freestone & McGoldrick, 2007). Guided by principles including sustainability, equitable labor practices, animal welfare, human rights, and corporate transparency, ethical consumers seek goods that embody minimal environmental impact (White et al., 2019), just labor compensation (Littrell & Dickson, 1999), and conscientious treatment of animals (Harper & Makatouni, 2002). Wettstein et al. (2019) argue that ethical labor practices foreground transparency, which serves as a metanorm, enabling consumers to navigate markets with an orientation toward justice and equity. In this framework, transparency operates as a guiding principle, urging clarity on the ethical footprint of consumer choices (New, 2015).

*Negative consumption,* or boycotting, represents a form of ethical abstention or moral resistance wherein consumers purposefully eschew products linked to unethical practices (Megicks et al., 2008; Kam & Deichert, 2020). This modality embodies the "power of absence," wherein boycotting becomes a mechanism for corporate accountability, compelling companies to amend practices or risk reputational and financial detriment. Empirical studies attest to the influence of this form of consumption in addressing unsustainable practices and environmental degradation (Dauvergne, 2018), illustrating its role in pressuring corporations toward ethical recalibration.

*Discursive consumption*, centers on communicative and advocacy-driven engagement, where consumers aspire to shape social norms and ethical standards beyond the act of purchase. Through online platforms, social media, and public forums, consumers engage in dialogue that amplifies values such as sustainability and fair trade (Berry & McEachern, 2005). The practices of brands like Patagonia and Ben & Jerry's exemplify "brand activism," wherein products are explicitly aligned with political and social ideals (Dubuisson-Quellier, 2013; Vredenburg et al., 2020). Such alignment fosters corporate social responsibility and catalyzes ethical market transformation by embedding moral values within the marketplace's fabric.

## The ethical price premium as moral capital

Stiefenhofer's work (2019), and Stiefenhofer and Zhang's research (2022) explore the interrelation between ethical pricing structures and consumer preference, suggesting that the premium paid for ethically produced goods constitutes a form of *moral capital*—a deliberate investment by consumers aligning with their ethical commitments. This readiness to incur additional costs for products embodying values like environmental stewardship and fair labor practices signals a departure from traditional economic rationality, leaning instead toward moral deliberation, where the ethical price premium signifies both social distinction and deeply held values in post-materialist societies.

Recent scholarship on *consumers' willingness* to pay (WtP) for ethically sourced goods corroborates this trend, evidencing a prioritization of ethical, social, and environmental values in consumer choices. Research by Fernández-Ferrín et al. (2023) and De Pelsmacker et al. (2005) illustrates how consumers frequently experience positive moral affect—such as pride—when purchasing fair trade or sustainable products, reinforcing an elevated WtP. Mai and Hoffmann (2015) further reveal that consumers oriented toward health and ethical standards perceive greater intrinsic value in sustainable products, thus justifying a premium. Additionally, Trudel and Cotte (2009) show that ethical practices elicit consumer rewards, while ethical lapses incur penalties, positioning ethical consumption as a mechanism of corporate accountability. Auger et al. (2010) confirm the cross-cultural significance of these dynamics, underscoring ethical concerns as a powerful influence on consumer behavior across diverse markets.

Empirical evidence particularly underscores this trend among younger demographics. Gomes et al. (2023) report that Millennials and Gen Z express marked preferences for sustainably sourced and organic goods, often willing to pay a premium for them. This *ethical inclination* permeates sectors such as fashion, where consumers favor ethically produced apparel and justify higher prices for responsible production practices (Davies et al., 2021). Similarly, in the circular economy, products designed for reuse and recycling command increased prices, emblematic of an emerging ethic centered on environmental responsibility (Sarkis, 2023). Collectively, this body of research illustrates a paradigm shift in consumer behavior, where ethical consumption shapes market dynamics, fosters corporate responsibility, and empowers consumers to enact and reinforce their values through economic choices.

## A Kantian theory of economic decision making

Economists have crafted a highly influential model of human decision-making, embodied in the metaphor of the *homo economicus* (Debreu, 1959) [2]. This construct is defined by traits fundamental to neoclassical economics, presenting a theoretical individual who is fully *rational*, making decisions aimed at maximizing personal utility or profit through logical reasoning. Governed by *self-interest*, the *homo economicus* seeks personal gain without regard for the welfare of others, unless such concern aligns with their own advantage. The model presupposes *perfect information*, assuming that individuals possess complete knowledge of all relevant factors, thereby enabling them to make optimal, calculated choices. At its core, the *homo economicus* is driven by utility maximization through material wealth and personal satisfaction, yet operates within the limits of *wealth constraints*. While this abstraction has proven valuable for market analysis within a materialist context, it has been critiqued for its reductionist view of human behavior, failing to account for the complexities of ethical considerations that emerge in both materialistic and post-materialistic contexts. Ethical considerations—such as environmental stewardship, the rights and dignity of labor, and the imperatives of corporate responsibility—frequently extend beyond the bounds of narrow self-interest, thereby challenging the reductionist assumptions of the homo economicus model, which prioritizes individual utility over the pursuit of collective welfare. These moral imperatives invoke principles of justice, solidarity, and responsibility, prompting actions motivated by duties to others and to the broader social and ecological order, rather than by mere personal gain.

This section critically examines the intersection of ethics and the neoclassical economic model, with a particular focus on the model's ability to incorporate ethical considerations. To ground this analysis, this paper limits its inquiry to *Kantian ethics*, with a specific emphasis on Kant's concept of *duty*. In *Kantian Ethics and Economics*, Mark D. White articulates a significant challenge to the neoclassical framework, positing that Kantian ethics disrupts the conventional economic model, where consumer behavior is predominantly understood as being guided by preferences and the pursuit of utility maximization (White, 2011). White's argument hinges on the Kantian distinction between *preferences* and *duty*, wherein preferences represent contingent desires or inclinations, while duty is an expression of moral law grounded in reason. According to White, Kantian ethics posits that moral duty can, and often must, supersede mere preferences; individuals are not morally free to act purely on subjective desires but are instead bound by objective duties that arise from the *categorical imperative*. Thus, in contrast to the neoclassical view, where agents are seen as utility maximizers whose preferences dictate their choices, a Kantian approach according to White suggests that *ethical constraints*—emanating from moral duty—must inform and sometimes limit economic decisions. This introduces a normative dimension that challenges the value-neutral assumptions of neoclassical economics, calling for a deeper consideration of moral imperatives alongside the pursuit of utility.

Kantian ethics offers a compelling approach to addressing moral issues in economics by emphasizing autonomy, dignity, and duty. Bowie (1999) argues that businesses must treat individuals as ends in themselves, not merely as means to profit. This principle is echoed by Hsieh (2004), who applies Kantian ethics to corporate responsibility, asserting that companies have both perfect and imperfect duties to their stakeholders, including employees and consumers. Arnold and Bowie (2003) extend this idea to labor practices, using Kantian principles to critique sweatshops by emphasizing respect for workers' dignity.

Similarly, Heath (2008) highlights that Kantian ethics, grounded in duty, provides a framework for moral motivation in business, challenging the self-interested assumptions of the homo economicus model. These perspectives reinforce the importance of embedding ethical principles into economic decision-making.

Subsequently, this study examines the extent to which the core principles of the neoclassical economic model must be reconciled or modified when viewed through the lens of Kantian ethics. Specifically, this research paper investigates how the model's emphasis on value-neutral analysis—centered on rational choice and utility maximization—conflicts with the normative requirements of Kantian ethics, particularly its focus on duty and moral law. The key question is whether incorporating Kantian ethical imperatives necessitates a fundamental revision of the neoclassical framework, or if these moral considerations can coexist with its descriptive approach without undermining its analytical foundations. To explore this, this paper turns to White's Kantian model of decision making (White, 2011).

## White's Kantian model of decision making

The analysis centers on Kant's concept of duty and its pivotal role in White's Kantian-economic model of decision making (2011). At the heart of this model, White introduces three ranking axioms that define how duties can be "seamlessly" incorporated into "preferences" and "constraints", shaping the decision-making process in a principled manner.

> Axiom I: *Ranking imperfect duties against inclination and other imperfect duties.*
> Axiom II: *Ranking perfect duties against each other.*
> Axiom III: Ranking perfect duty against imperfect duty.

White's model artfully integrates Kant's concept of duty into the prevailing framework of preferences and constraints, offering a reimagined approach to economic decision-making. This incorporation of duty is substantiated by a clarification that preferences, within the context of modern economic theory, need not imply alignment with subjective mental states such as happiness or pleasure. Rather, preferences signify an *ordered hierarchy of possible states of the world* over which the agent exercises a degree of influence, thus reflecting a rational structuring of values beyond mere inclination (White, 2011, p. 42).

To clarify the concept of "preference", which is mathematically represented by a preference relation (≥), it is crucial that both the *consumption space X* and the goods $x$ within it are precisely defined [3]. The consumption space, encompassing all possible combinations of goods available to the consumer, forms the foundation of the decision-making process. In the neoclassical model, the consumption space is denoted as $X = \{ x \in \mathbb{R}^n_+ : x_i \geq 0 \ \forall \ i = 1, 2, \ldots, n \}$, where $x = (x_1, x_2, \ldots, x_n)$ is a consumption bundle and each $x_i$ for $i = 1, 2, \ldots, n$ is defined by its *physical*, *temporal*, and *spatial* attributes, as well as its non-negative *quantity* (Debreu, 1959).

In White's model, both the consumption space, which represents the full range of economic choices available to the agent and the *duty space*, which should delineate the scope of duties to be ranked and followed, remain as implicit assumptions. The absence of explicit definitions for these spaces leaves the interaction between economic choices and moral duties insufficiently specified, weakening the model's theoretical clarity and its capacity to integrate Kantian ethics into decision-making frameworks. This lack of formalism leaves the model underdeveloped, as it obscures the relationship between duties and consumption. Without clearly specifying the domains in which choices occur and the characteristics of the goods and duties involved, the concept of a preference relation "≥" lacks its foundational structure. Only with a precise delineation of these spaces can preferences be understood as an ordered ranking of goods and duties, reflecting how agents evaluate different bundles based on their attributes, independent of subjective states like happiness or pleasure. Consequently, the lack of a rigorous formulation for both spaces undermines the model's ability to authentically integrate Kantian notions of duty within the framework of economic decision-making. This paper aims to address this limitation directly.

A preference relation satisfying *reflexivity*, *completeness*, and *transitivity* offers a coherent framework for rationally ordering goods, illustrating how individuals systematically rank outcomes. Arrow and Debreu's influential formalization (1954), inspired by the *Bourbaki* group's pursuit of *formalism* and *axiomatic* rigor, extended these principles into economics, providing a precise model for understanding market behavior and equilibrium [4]. Following this tradition, Arrow and Debreu axiomatized preference relations using binary relations to logically structure choices, echoing Bourbaki's emphasis on reducing concepts to their fundamental logical cores. This approach enabled precise and consistent modeling of economic behavior. By *prioritizing internal consistency and logical coherence,* Arrow and Debreu infused the mathematical rigor of the Bourbaki tradition into economics, transforming it into a formal, axiomatic discipline. In a similar vein, White asserts that:

> *"So, in theory, preferences can also be based on Kantian duty as derived from the categorical imperative; the agent could rank states of the world in which she performed her duty higher than ones in which she did not, without implying any 'desire' for the former or later. Furthermore, if the agent can rank some duties higher than other duties as well as her 'normal' preferences, and can do so completely and transitively, then these duty-based 'preferences' can be included in an ordinal utility function"* (White, 2011, p.42).

**White's conjecture:** If duties can be ranked—setting aside the specific nature of duty for the moment— and satisfy completeness and transitivity, then there exists an "ethical" utility function.

The precise ontological and epistemological status of White's conjecture remains shrouded in ambiguity: it is uncertain whether the claim pertains to the rationality of moral agents or to the formal existence of an ethical utility function. Demonstrations of rationality traditionally invoke the reflexivity axiom as a foundational premise, whereas the derivation of utility functions typically necessitates more robust formal conditions. The absence of a rigorous formalization of the consumption and duty spaces compounds this uncertainty, rendering the conjecture's conceptual boundaries and practical implications indeterminate. This lack of definitional clarity gestures toward a deeper philosophical need for a more rigorous and coherent articulation of the theoretical framework in which such propositions are embedded, ensuring their intelligibility and applicability.

Moreover, White does not explicitly require *duty-based preferences* to adhere to the additional properties of *monotonicity, continuity*, or *convexity* that are typically required to establish existence of a *continuous utility function* [5]. While the requirements of completeness and transitivity for duty-based preferences ensure that all duty-bundles can be systematically compared and consistently ranked—thereby avoiding cyclical contradictions in choice—further conditions are necessary to render these preferences fully coherent and meaningful [6]. To clarify these additional requirements, this research examines the properties of duty in greater depth.

The distinction between *perfect* and *imperfect duties*, derived from Kantian ethics, provides a valuable framework for analyzing both mandatory and discretionary behaviors in economic contexts. This distinction is based on *Kant's categorical imperative*, specifically the *Universal Law formulation.*

*Perfect duties* emerge when a maxim, if universalized, leads to a logical contradiction, such as the prohibition against lying, and must be followed unconditionally (Kant, 1785, 4:421-424). *Imperfect duties,* like benevolence, arise when universalization doesn't cause a contradiction but cannot be willed universally (Kant, 1785, 4:423-424).

According to Kant, *perfect duties* [7] are strict, binding obligations that must be fulfilled in all circumstances. Examples include paying taxes, honoring contracts, and refraining from illegal or unethical activities. These duties are absolute and non-negotiable, often reinforced by legal frameworks to maintain trust and fairness in economic systems. Their qualitative nature is evident in their unyielding application, distinguishing them from discretionary actions and underscoring their role in upholding moral and societal order. *Imperfect duties* are moral obligations that allow for flexibility regarding how, when, and to what extent they are fulfilled. Unlike perfect duties, which must always be met, imperfect duties, such as acts of charity or self-improvement, permit discretion in their execution. Individuals can choose the timing and manner of fulfilling these duties, reflecting the fact that they are not absolute, but still contribute to moral good and societal welfare. Economic examples include corporate social responsibility, charitable donations, and sustainable business practices. While these actions are not legally required, they reflect a moral commitment to contributing to societal welfare or the environment. The extent to which individuals or businesses engage in such practices depends on their discretion, resources, and values.

White acknowledges that Kant's distinction between perfect and imperfect duties presents a profound challenge for integrating *duty-based preferences* into traditional economic models. To address this, White deviates from standard frameworks by treating perfect duties not as elements of the preference ordering but as *external constraints on behavior.* In this interpretation, imperfect duties inform flexible, subjective preferences, while perfect duties remain absolute moral imperatives, separate from preference optimization. This approach fundamentally reshapes the integration of moral obligations into economic decision-making, moving beyond strict utility maximization to incorporate ethical nuance.

> *"Plainly, perfect duties cannot be included among preferences, since they take precedence over inclinations and cannot be treaded off amongst them if performing them becomes too expensive. Rather, perfect duties constrain the pursuit of inclinations, spelling out the means we may not use to achieve our ends"* (White, 2011, p.42).

This conclusion stems from Kant's *Formula of Respect*, which asserts that one must never treat others—or oneself—merely as a means to an end, but always also as ends in themselves (Kant, 1785). This moral principle forbids actions that exploit or instrumentalize individuals for personal gain without acknowledging their intrinsic worth and autonomy, ensuring that all persons are treated with the dignity and respect due to rational agents. White then concludes that a necessary condition for modelling perfect duties is to include them as a constraint:

> *"With this understanding, we can model perfect duties as constraints, in the same way that budget constraints limit customer's spending"* (White, 2011, p.42).

In *Groundwork of the Metaphysics of Morals*, Kant (1785) offers an example of a shopkeeper who is honest with his customers because it is good for business. The shopkeeper refrains from cheating not out of respect for the customer's rights but to maintain his reputation and profits. Kant argues that while the action (being honest) aligns with duty, it has no true moral worth because it is driven by self-interest rather than a genuine respect for the customer as a rational being deserving of fair treatment (Kant, 1785, p. 11). White does not explicitly define the duty space, making it challenging to infer a clear concept of a constraint that would delineate a "duty-feasibility sub-space" within the broader duty framework [8]. This lack of specification leaves the notion of operating within a well-defined duty space ambiguous and open to interpretation, an issue that this paper will address in great detail.

Other examples of perfect duties in economics reflect absolute moral obligations that must be followed without exception, aligning with Kantian ethics. For example, businesses have a perfect duty to act with honesty in transactions, such as providing truthful information about products and disclosing risks to clients, regardless of potential profits. Likewise, the duty to fulfill contracts is binding, requiring parties to honor agreements even if circumstances make it less profitable. Employers also have a perfect duty to pay fair wages, ensuring that workers are compensated as agreed, without exploitation. Additionally, avoiding fraud and corruption in business dealings, such as refusing bribes or refraining from insider trading, is an inviolable duty. Finally, respecting property rights by not infringing on intellectual or physical property, even when doing so would offer competitive or financial advantage, remains a fundamental obligation. These duties underscore the ethical foundation of economic behavior, where actions must respect the rights and dignity of others, irrespective of personal gain. White argues that perfect duties are implicitly assumed in the neoclassical model and concludes that

*"Ironically, the nature of perfect duty seems to be the tallest hurdle to overcome when integrating Kantian ethics into the economic model of choice, but in a way, it has been there all along"* (White, 2011, p.42).

The Arrow-Debreu model (1954) implicitly integrates certain perfect duties within its idealized structure, though these remain unarticulated. It presupposes that agents will honor contracts, provide complete and truthful information, refrain from generating harmful externalities, and engage equitably in transactions. These underlying assumptions gesture toward a Kantian moral landscape, where agents fulfill ethical obligations essential to the model's coherence. However, this static moral backdrop assumes fixed ethical foundations without addressing the origins or evolution of these standards, limiting the model's engagement with moral development and transformation within economic systems.

White proposes an explicit incorporation of perfect duties into a Kantian ethical model, drawing an analogy to the embedding of budget constraints within neoclassical economic frameworks; however, this analogy proves to be conceptually strained. Modelling duties, particularly perfect duties, as analogous to budget constraints exposes significant foundational challenges. While budget constraints are continuous, measurable, and situated within a rigorously defined economic space $X \subseteq \mathbb{R}_+^n$, moral duties resist such formalization. Ethical obligations are often binary and context-dependent, lacking the divisibility and continuity that characterize economic spaces. Moreover, duties are inherently qualitative, frequently embodying moral absolutes that stand in stark contrast to the incremental adjustments central to economic behaviour. Despite White's attempt to reconcile duties with an economic framework, the binary and non-divisible nature of moral obligations remains fundamentally incompatible with the continuous structures of such models. Nevertheless, this study will explore how topological manifolds may provide a meaningful framework for constructing a duty-space, offering a potential bridge between ethical and economic formalizations.

In light of these distinctions, this paper now turns to an analysis of the ranking axioms in White's model, focusing initially on Axioms II and III before revisiting Axiom I to propose refinements that might better integrate Kantian duties within the model's structure.

# Axiom II: Ranking perfect duties against each other

In Kantian ethics, perfect duties are characterized by their universality, absolute nature, and non-contradictory status. The mathematical properties of perfect duties must reflect their logical consistency, universality, and unconditional obligations. These duties form the cornerstone of Kant's deontological framework, directing moral actions through rational deliberation and strict adherence to moral law. By examining these properties, this study can gain valuable insights into the role of perfect duties within the broader context of Kantian ethics, showcasing the inherent structure and rigor of his moral philosophy. This framework not only underscores the significance of duty in ethical decision-making but also emphasizes the critical function of reason in discerning moral imperatives.

While perfect duties in Kantian ethics resist representation through continuous preferences due to their absolute, non-negotiable, and non-quantifiable nature, they can, in principle, be integrated into preference frameworks by employing discrete preference orderings that honour moral imperatives. Consequently, representing perfect duties requires a departure from traditional complete metric space models toward frameworks that respect the categorical nature of these ethical obligations, thereby capturing the full complexity of moral duties.

White argues that perfect duties should be introduced as constraints similarly to budget constraints in the neoclassical model. Together with the second axiom White (2011) suggests some ranking of constraints.

> *"In a sense, perfect duties shrink the 'action space' available to a moral agent: they foreclose all the actions that involve deception, coercion, and so forth".*

This paper takes a significant departure from this perspective. Having demonstrated that the space of real numbers may not be ideal for modeling perfect duties, this analysis turns its focus to a *discrete space*, where constraints acquire a new interpretation. This proposes that the universe of perfect duties is most accurately represented as a discrete topological space. In this framework, every subset is open, allowing each point—symbolizing a distinct duty—to remain entirely isolated from the others. This inherent separation renders the discrete space highly flexible and straightforward, providing an effective foundation for conceptualizing perfect duties.

Utilizing the idea of a discrete topological space offers a counterpoint to White's suggestion that imperfect duties act as constraints reducing the action space. In a discrete topological framework, each perfect duty is represented as an isolated, independent element, which allows for flexibility and autonomy in ethical decision-making. This suggests that imperfect duties, rather than functioning as rigid constraints, may instead operate within an open and *non-hierarchical space,* where they guide behavior without imposing strict limitations. Thus, a discrete topology challenges White's perspective by indicating that imperfect duties do not necessarily impose rigid constraints on the "action space" but can coexist within a more flexible and adaptable ethical framework, allowing for greater fluidity and context-dependence in moral decision-making. This paper proceeds by constructing the topology of the perfect duty space in conjunction with the economic space, illuminating the interplay between ethical obligations and economic constraints.

Kant's three formulations of the *Categorical Imperative—The Formula of Universal Law, The Formula of Humanity*, and *The Formula of Autonomy—* serve as a rigorous framework for determining moral duties by emphasizing universality, respect for human dignity, and rational self-legislation (Kant, 1785,1797). These formulas guide moral reasoning by requiring that actions be universalizable, respect others as ends in themselves, and align with principles suitable for a community of rational agents. A set of perfect duties Y arises when a maxim fails the test of universalizability or undermines human dignity, creating absolute prohibitions, such as the duty not to lie or commit suicide. A set of imperfect duties $E$ emerges when actions promote moral ideals without contradiction but allow for flexibility in their fulfillment, such as the duty to aid others or cultivate one's talents. Together, these categorical imperatives help construct a structured ethical system I, balancing necessary obligations with aspirational virtues.

$$I \rightarrow \{Y, E\} \tag{1}$$

Let Y be a *discrete topological* space of dimension $m$, where Y=$\{y_1, y_2, ..., y_m\}$ represents the qualitative properties of perfect duties. The set $E$ will be introduced later. Let $\mathbb{R}^n_+$ represent the real space of dimension $n$ with the *standard topology* representing the economic space in which commodities are traded. The goal is to construct the *product topology* on Y×$\mathbb{R}^n_+$ thereby integrating the framework of perfect duties with the dynamics of economic activities. The discrete topology is characterized by a unique set of axioms that distinguish it from other topological spaces. First, the empty set $\emptyset$ and the entire set $Y$ must be open, which holds in any topological space: $\emptyset, Y \in T$, where $T$ is the collection of open sets. In the discrete topology, every subset of $Y$ is open, i.e., $T = P(Y)$, the power set of $X$. This leads to the second axiom, stating that arbitrary unions of open sets are open: if $\{U_\alpha\}_{\alpha \in I}$ is a family of open sets, then $\bigcup_{\alpha \in I} U_\alpha \in T$, where $I$ is a finite countable set. Finally, the third axiom asserts that finite intersections of open sets are open: if $U_1, U_2, \ldots, U_n \in T$, then $U_1 \cap U_2 \cap \ldots \cap U_n \in T$. Given that every subset of $Y$ is open, these axioms are trivially satisfied, making the discrete topology the finest possible topology, where $T$ contains all subsets of $Y$. In the discrete topology on a set $Y$, hence, every subset of $Y$ is considered an open set, including the empty set and Y itself. Every subset of $Y$ is open, and in particular, the singleton sets $\{y_i\}$ for $i = 1, 2, \ldots, m$ are open.

Consider $\mathbb{R}_+^n$ as the n-dimensional space equipped with the standard topology, where open sets are unions of open n-balls. The product topology on $Y \times \mathbb{R}_+^n$ is generated by the basis of sets of the form: $U \times V$, where $U \subseteq Y$ is open and $V \subseteq \mathbb{R}^n$ is open. Since Y is discrete, any subset $U \subseteq Y$ is open, and open sets in $\mathbb{R}^n$ are typically open n-balls or unions of such balls. Since Y is discrete, any subset of Y is open, meaning each point in Y can be treated individually. The product topology will inherit this property, and the space $Y \times \mathbb{R}_+^n$ can be visualized as having m "copies" of $\mathbb{R}_+^n$. The real space $\mathbb{R}_+^n$ retains its continuous structure, where open sets are n-balls or unions of them.

We can now identify the structure of the space $Y \times \mathbb{R}_+^n$. For each point $y_i \in Y$, the "slice" $\{y_i\} \times \mathbb{R}_+^n$ behaves like a copy of $\mathbb{R}_+^n$. Thus, $Y \times \mathbb{R}_+^n$ can be viewed as $m$ disjoint "layers" or copies of $\mathbb{R}_+^n$, one for each $y_i \in Y$. Open sets in $Y \times \mathbb{R}_+^n$ will be unions of the form: $\bigcup_{(i)} (\{y_i\} \times V_i)$ where $V_i \subseteq \mathbb{R}_+^n$ is open. Each $\{y_i\} \times V_i$ corresponds to an open set in the real space $\mathbb{R}_+^n$ associated with the point $y_i \in Y$. The Space $Y \times \mathbb{R}_+^n$ has now a natural interpretation.

In this framework, the discrete space $Y$ represents the qualitative properties of perfect duties, where each element signifies a distinct ethical obligation, such as "not to lie" or "to respect others' rights." The real space $\mathbb{R}_+^n$, on the other hand, captures the economic dimension of commodity trading, with each point representing a specific allocation of goods. By combining these spaces within a *product topology*, the model facilitates the examination of how moral imperatives influence economic decisions. Each slice $\{y_i\} \times \mathbb{R}_+^n$ represents a distinct scenario, where a specific perfect duty or combination of perfect duties is in effect, corresponding to a unique economic configuration. This structure enables a clear analysis of the interplay between ethical obligations and economic outcomes.

The framework described aligns with a *fiber bundle approach* because it systematically organizes the interaction between ethical duties and economic configurations. In a fiber bundle, the *base space* represents an underlying set of contexts or parameters—in this case, the space $Y$ of ethical duties. Each ethical duty bundle $y_i \in Y$ corresponds to a *fiber*, which consists of all possible economic allocations in $\mathbb{R}_+^n$ that are considered under that specific duty bundle. The total space $Y \times \mathbb{R}_+^n$ encompasses all combinations of ethical obligations $y_i \in Y$ and economic allocations $x \in \mathbb{R}_+^n$ while the *projection map*

$$\pi: Y \times \mathbb{R}_+^n \to Y \tag{2}$$

assigns each combination to its respective ethical duty. This structure enables the analysis of how different ethical contexts influence the range of permissible economic decisions, highlighting the relationship between duties and outcomes. By using a fiber bundle framework, this analysis achieves a clear separation between ethical contexts and economic choices, allowing for a more precise understanding of how changes in ethical obligations affect consumer behavior and economic configurations.

# Axiom III: Ranking perfect duty against imperfect duty

In Kantian ethics, *imperfect duties* are established through the application of the *categorical imperative,* specifically in its second and third formulations. These duties are moral obligations that, unlike perfect duties, do not require specific actions in every situation but instead allow for discretion in how and when they are fulfilled. Imperfect duties are rooted in Kant's broader ethical framework, which emphasizes the respect for rational agents and the promotion of moral ends.

In the context of sustainable economics, imperfect duties arise from the moral obligation to promote long-term environmental and societal well-being, even though the specifics of action are left to individual discretion. For example, the duty to reduce environmental impact (e.g., through minimizing waste or lowering carbon emissions) is an imperfect duty, as businesses and individuals are morally encouraged to contribute to sustainability but are not required to act in the same way or to the same degree in every instance. Another imperfect duty in this context is the obligation to support fair labor practices, which entails improving working conditions and ensuring fair wages, but allows flexibility in how companies implement these practices. These duties are established through the broader moral principles of *beneficence* and *justice*, urging economic agents to contribute to societal and environmental welfare without rigidly prescribing specific actions. In Kantian terms, these duties promote the good of others while respecting the autonomy of individuals and organizations in how they fulfill them.

> *"If you planned to donate $1,000 to charity, and then realized that you had forgotten a debt of $500 that you have to repay, you can always reduce your* *donation without forgoing it altogether. More generally, you still hold the end of charitable giving, but you decide that planned, specific execution of the corresponding duty must be compromised to some extend"* (White, 2011, p.49) [9].

We have already constructed the product topology on $Y \times \mathbb{R}_+^n$ thereby integrating the framework of perfect duties with the dynamics of economic activities in the consumption space. However, the next step is to establish a foundation for modeling imperfect duties. This requires extending the consumption space. Let's consider the extended consumption space $D := X \times E = \{ x \in \mathbb{R}_+^n, e \in \mathbb{R}_+^l : x_i, e_j \geq 0 \ \forall \ i = 1, 2, \ldots, n, \forall \ j = 1, 2, \ldots, l \}$, where $x = (x_1, x_2, \ldots, x_n)$ is a consumption bundle and each $x_i$ for $i = 1, 2, \ldots, n$ is defined by its physical, temporal, and spatial attributes, as well as its non-negative quantity, and $e = (e_1, e_2, \ldots, e_l)$ is an a imperfect duty-bundle and each $e_j$ represents an imperfect duty satisfying the same properties. The space $E$ is generate by model (1). In this model consumer preferences extend beyond mere material consumption to encompass ethical choices, reflecting Kantian principles of moral duty in a cross product. Consumers are portrayed as making decisions not only based on physical consumption $x$, which delivers conventional utility through comfort or pleasure from goods, but also on moral consumption $e$, which provides a distinct form of utility derived from fulfilling imperfect duties. These ethical actions, such as helping others or self-improvement, contribute to a sense of moral worth, aligning with Kant's view that ethical obligations play a crucial role in human decision-making and well-being.

In this $l$-dimensional space $E$, each dimension represents an imperfect duty, such as charitable giving, self-improvement, or aiding others. Preferences arise from the varying degrees to which individuals fulfill these duties, with completeness ensuring that any two combinations of duty fulfillment can be compared, and transitivity maintaining logical consistency in these judgments. This l-dimensional real space of distinct imperfect duties enables a structured method for ranking moral actions, allowing individuals to assess and prioritize their duties systematically. Much like in economic models, this framework facilitates thoughtful consideration of trade-offs, creating a balanced yet flexible approach to fulfilling moral obligations.

The product space $Y \times D = Y \times (X \times E)$ incorporates both the discrete space of perfect duties $Y$ and the $n + l$ dimensional economic-ethical consumption space $D$, which includes physical consumption $X$ and moral consumption $E$. The *product topology* on $Y \times D$ is defined as follows: Open sets in the product space $Y \times D$ are generated by taking products of open sets from $Y$ and open sets from $D$. Since $Y$ has the discrete topology, any set of the form $\{y_i\} \times U$, where $U$ is an open set in $D$, is an open set in $Y \times D$. Open sets in $D = X \times E$ are formed as products of open sets in $X$ and open sets in $E$, where the standard topology applies.

In Kantian ethics, the discrete topological space $Y$ symbolizes the realm of categorical moral imperatives, corresponding to *perfect duties*. These are absolute, non-contingent obligations that apply universally, admitting no exceptions. Each point $y_i \in Y$ can be understood as representing a specific perfect duty, such as the prohibition against lying or the obligation to avoid harm, duties that must be adhered to in all circumstances. On the other hand, the space $D = X \times E$ integrates both economic activities and the realm of *imperfect duties*. $X$ represents the material and empirical dimension of consumption and economic decision-making, while $E$ encompasses the ethical domain of imperfect duties. Imperfect duties, in Kant's framework, are moral obligations that, while binding, allow for discretion in their execution and do not require specific actions at every moment. These duties—such as promoting personal development or engaging in acts of charity—guide moral agents toward virtuous conduct, providing ethical direction that is sensitive to context and circumstance, yet not as rigidly prescriptive as perfect duties.

The *product topology* $Y \times D$, where $Y$ is a discrete space and $D = X \times E$ represents the consumption-ethical space, can be viewed as a specific case of a *trivial fiber bundle*. In this case, the *base space* $Y$ is a discrete space, and the *fiber space* $D$ remains constant over each point in the base space. In the product space $Y \times D$, $Y$ plays the role of the *base space* and $D$ acts as the *fiber space*, which is the same for every point in $Y$. The *projection map* $\pi: Y \times D \to Y$ sends each point $(y, d) \in Y \times D$ to its corresponding point $y \in Y$.

This product topology $Y \times D$ is an example of a trivial fiber bundle, where the base space $Y$ has the same fiber space attached to each point. This is a special case of a fiber bundle, where the total space is simply the Cartesian product of the base and fiber spaces. More complex fiber bundles can have fibers that vary over the base space, leading to richer topological structures that cannot be represented as simple product spaces.

Let's consider an example of a *non-trivial fiber bundle* with $Y$ and $D$. Suppose $Y = \{y_1, y_1\}$ is a discrete space with two points. Instead of having a trivial product $Y \times D$, where the same fiber $D$ is associated with both $y_1$ and $y_2$, let's introduce a fiber bundle where the fiber spaces differ at each point. Let's assume $D$ consists of two different spaces of consumption and ethical choices:

i.  At $y_1$: The fiber $D_{y_1}$ is $D_{y_1} = X_1 \times E_1$ , where $X_1 \subseteq \mathbb{R}^n_+$ represents a specific space of economic activities, and $E_1 \subseteq \mathbb{R}^l_+$ represents a space of imperfect duties that are relevant only at $y_1 \in Y$.

ii.  At $y_2$: The fiber $D_{y_2}$ is $D_{y_2} = X_2 \times E_2$ , where $X_2 \subseteq \mathbb{R}^n_+$ represents a different space of economic activities (perhaps with different constraints or opportunities) and $E_2 \subseteq \mathbb{R}^l_+$ represents a different set of imperfect duties, $y_2 \in Y$.

Here, the fiber spaces $D_{y_1}$ and $D_{y_2}$ differ, reflecting the idea that the nature of economic and moral decisions may change based on the underlying qualitative properties represented by $Y$. The total space $Y \times D$ is a more nuanced structure that varies with the base space $Y$. This nuanced structure aligns with Kant's distinction between duties: while perfect duties might be universal and constant, imperfect duties can vary depending on particular circumstances, illustrating the interplay between unchanging moral principles and context-sensitive moral action. Figure 1 shows equilibrium allocations $x_j = (x_j, e_j)$ in each fiber $D_{y_j}$ associated with a bundle in the base space $y_j \in Y$. The projection mapping allows to trace exchange between economic agents in different ethical contexts starting from some initial endowment $\omega_1$ [10].

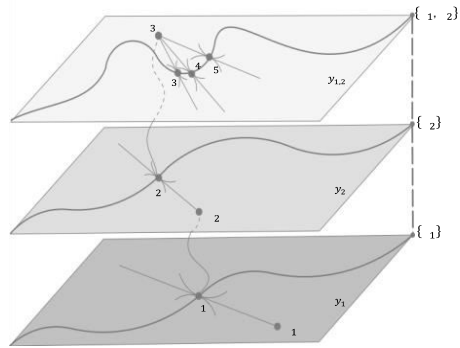

*Figure 1: Example of a Fiber Bundle Model*

The fiber bundle approach offers a further advantage: it retains both foundational assumptions of Inglehart's model of intergenerational value change, contrasting with Kant's static view, by allowing for a time-parameterized evolution in the configuration of perfect duty bundles. The first assumption posits that scarcity shapes value priorities, so that in times of economic or existential insecurity, individuals prioritize materialist values, such as economic stability and physical safety. The second assumption suggests that as societies attain higher levels of economic and physical security, subsequent generations gravitate toward post-materialist values, emphasizing self-expression, autonomy, and quality of life. This dynamic progression will be examined in detail in the following section.

## Intergenerational shift in values: the case of slavery

Inglehart's theory posits that economic and existential conditions shape generational values. In early generations, where economic survival depended on systems like slavery, materialist values predominated. Slavery was justified as necessary for maintaining economic stability, property rights, and social order (Inglehart, 1977; Genovese, 1974). Social hierarchies and racial divisions reinforced these beliefs, embedding slavery within the societal fabric (Berlin, 1998).

However, as industrialization and economic diversification advanced, existential security improved. Younger generations embraced post-materialist values, prioritizing human rights, equality, and justice over economic concerns. The moral imperative to respect individual dignity catalyzed abolitionist movements, leading to the eventual recognition of slavery as a profound moral wrong (Inglehart, 1990; Hunt, 2007).

The fiber bundle approach formalizes this evolution, illustrating how economic and moral decisions about slavery evolved through shifting historical contexts. For example: $y_1$ might represent a period where slavery is economically essential and morally accepted. $y_2$ might represent a later period where moral and economic attitudes have shifted, and abolition movements begin to gain traction, and $y_3$ might represent a post-abolition period, where slavery is universally condemned and illegal. The fiber spaces $D_{y_1}$, $D_{y_2}$, $D_{y_3}$ represent the range of economic and moral decisions available in each context.

At $y_1$, the era endorsing slavery, the fiber space $D_{y_1} = X_1 \times E_1$ depicts a social and economic order where slavery is regarded as morally acceptable and economically vital. Economic decisions within $X_1$ are directed toward profit maximization through the exploitation of enslaved labor, while the moral framework of $E_1$ reinforces a worldview that legitimizes and sustains this institution. At $y_2$, a transitional time of shifting perspectives, the fiber space $D_{y_2} = X_2 \times E_2$ captures a society beginning to grapple with the moral and economic implications of slavery. Economic choices within $X_2$ start to include alternatives to slave labor, influenced by industrial advances or evolving production methods, while moral views in $E_2$ increasingly recognize the injustice of slavery and the emerging concept of human rights. Finally, at $y_3$, the post-abolition period, the fiber space $D_{y_3} = X_3 \times E_3$ reflects a society that has embraced fundamental change. Economic structures within $X_3$ are built upon free labor markets, and the moral framework $E_3$ universally condemns slavery as a profound violation of human dignity. Here, values of autonomy, equality, and respect for human dignity dominate, reshaping both the moral and economic landscape toward a more enlightened, post-materialist society.

The projection mapping in this fiber bundle framework models intergenerational value transitions, illustrating how societal values evolve across generations. Specifically, the projection map $\pi: D \to Y$ serves as a powerful representation of these global shifts, capturing the nuanced interplay between moral duties and economic decisions over time. This projection allows to see how values, including Kantian *perfect duties* (strict obligations, such as respecting others' rights) and *imperfect duties* (moral ideals, such as benevolence), adapt and interact with economic frameworks as society advances. Each point $y_i \in Y$ corresponds to a fiber space $D_{y_i}$ that reflects the distinct moral and economic configurations of that period.

In the case of slavery, it can be seen, at $y_1$, a paradigm in which material scarcity renders slavery seemingly necessary, a pragmatic response to limited resources. Yet as societies advance—reflected in the shifts to $y_2$ and ultimately $y_3$—fiber spaces reveal evolving economic structures that transcend this dependence on slavery. Concurrently, a moral progression unfolds, where new imperatives grounded in Kantian ethics emerge, prioritizing autonomy, dignity, and freedom as intrinsic human values.

At $y_3$, society embodies a moral landscape aligned with Kantian principles, treating individuals as ends in themselves rather than as means to an economic end. This stage marks a culmination of value evolution, resonating with Inglehart's theory: that changing material conditions foster shifts from survival-based values to those emphasizing self-expression and moral autonomy.

The fiber bundle framework thus elegantly captures this dynamic moral and economic evolution, offering a nuanced lens through which to understand the intertwined transformations of ethics and economic imperatives over time. It reveals not a static moral schema but an adaptive progression, as each $y_i$ crystallizes the particular moral and economic consciousness of its epoch.

We have demonstrated that the fiber bundle approach seamlessly unifies White's Kantian model of economic decision-making with Inglehart's theory of intergenerational value change, offering a structured framework that dynamically represents shifts in economic choices and moral values across varied contexts. Within this framework, the base space $Y$ represents different contexts—whether historical, generational, or ethical paradigms—each framing a distinct set of conditions that influence both economic and moral decisions. The fiber spaces $D_y$ encapsulate the specific economic choices and moral imperatives accessible within each context, while the projection map establishes a linkage between contexts and the decisions they inform, thereby charting the evolving landscape of values and economic behaviors over time.

This model captures both the universal aspects of morality, embodied in perfect duties, and the context-sensitive aspects of values, reflected through imperfect duties and the intergenerational progression of ideals. In this way, the fiber bundle approach provides a powerful and flexible generalization that transcends the limitations of either model alone. Notably, the fiber bundle approach operates independently of the ranking axiom central to White's model (2011), thus offering a broader and more adaptable framework for understanding the interdependence of ethical principles and economic realities across time and space.

# Axiom I: Ranking imperfect duties against inclination and other imperfect duties.

This paper has explored a specific approach to extending the consumption space to incorporate imperfect duties. By assuming that these duties conform to standard axioms of preference, the vector bundle framework partially addresses the integration of ethical dimensions within the economic model. Alternative frameworks warranting consideration include Lancaster's *New Consumer Theory* (1966), which reconfigures consumption analysis around the characteristics of goods rather than the goods themselves, thus potentially accommodating ethical attributes. Additionally, models with price-dependent ethical preferences, such as those advanced by Stiefenhofer (2021a, 2021b), offer further insight into the complexities of incorporating moral considerations within consumer choice.

*"Imperfect duties, even if they generally fit nicely among inclinations in the agent's preference ranking, must nonetheless be ordered somehow amongst them, as well as against each other"* (White, 2011. p. 47).

The fiber bundle approach offers a robust framework to unify and extend Inglehart's theory of intergenerational value change and White's Kantian model of economic decision-making, presenting a cohesive perspective on the evolution toward sustainable democracy. Inglehart's theory posits that economic development and security catalyse generational shifts toward post-materialist values, emphasizing environmental protection, human rights, and democratic ideals (Inglehart, 1990). Concurrently, White's Kantian model emphasizes the ethical imperatives within economic choices, where individuals aim to fulfil moral duties that respect human dignity and autonomy, beyond mere personal gain (White, 2011). By integrating these frameworks within a fiber bundle model, this analysis considers shifts in value-laden preferences, a dynamic often treated statically in existing literature. This model elucidates how evolving economic conditions, moral values, and political structures intertwine to foster democratic systems grounded in human dignity, equality, and justice.

This study posits that the projection mapping $\pi: Y \times D \to Y$, where $D = X \times E$, uncovers a deeper global interaction between shifting values and democratic development by integrating the domain of perfect duties $Y$ with the combined imperfect duty-economic space $D$. This mapping constructs a global framework in which individual decision-making is not isolated but is instead situated within a dynamic, interdependent ethical and economic context. The structure of the *projection mapping*

$$\pi: Y \times D \to Y \tag{2}$$

goes beyond a simple reflection of choices within various societal contexts. By linking individual choices, ethical imperatives, and economic structures, the mapping illuminates pathways through which personal actions align with broader ethical standards, shaping collective behaviours and supporting a foundation for sustainable democracy.

In this way, the mapping serves as a bridge between individual agency and collective ethics, showing how individual decisions informed by ethical duties contribute to societal well-being. This integrative framework encourages democratic sustainability by fostering an environment where choices are informed not solely by personal utility but by shared ethical commitments, balancing individual freedoms with responsibilities toward society's enduring health and equity.

In this setup, the fiber spaces $D_y = X_y \times E_y$ at each point $y \in Y$ contain both individual choices $X$ and the broader ethical-economic context $E_y$. Here, the choices within $X$ are influenced not only by societal ethics but also by the underlying economic and legal structures within $D_y$, which evolve as $Y$ transitions from one ethical framework to another (e.g., from slavery acceptance $y_1$ to abolition $y_2$ to democratic equality $y_3$). Consequently, the projection mapping can articulate how individual choices are filtered and shaped through the combined influence of ethical duties, societal norms, and economic realities.

The projection map $\pi$ highlights how shifts in ethical contexts in $Y$ directly influence both personal and systemic dimensions of decision-making [11]. For instance, under $y_1$, where slavery is accepted, the fiber $D_{y_1} = X_{y_1} \times E_{y_1}$ represents choices embedded within an economic structure of exploitation. Decisions within $X_{y_1}$ and $E_{y_1}$, therefore, align with this structure, reinforcing practices that sustain slavery. As the ethical context shifts to $y_2$, moving toward abolition, the economic context $E_{y_2}$ begins to incorporate labor rights and reject exploitative practices. The fiber $D_{y_2} = X_{y_2} \times E_{y_2}$ then supports individual decisions within $X_{y_2}$ that increasingly respect individual autonomy, human dignity, and fair labor practices. By the time the projection reaches $y_3$, democratic values of justice and equality are fully integrated, allowing $D_{y_3} = X_{y_3} \times E_{y_3}$ to reflect individual and economic choices that are harmonized with principles of inclusive democracy.

The projection mapping's structure reveals how individual choices in $X$ recursively influence the evolution of $E$, creating a feedback loop that shapes both personal and systemic aspects of democratic progress. As individuals make ethical and economic decisions that align with evolving ethical principles (such as abolitionist values in $y_2$ and democratic ideals in $y_3$), these choices reinforce and solidify the ethical-economic structures within $D$. This recursive interaction between $X$ and $E$, mediated by the progression through $Y$, shows that sustainable democracy is not merely the product of top-down ethical shifts but also the result of cumulative, ethically informed individual decisions [12].

By capturing the interplay between individual choices $(x, e) \in D$ and evolving ethical contexts $Y$, the projection mapping $\pi: Y \times (X \times E) \to Y$ provides a comprehensive model for understanding shifts in democracy. It reveals that democratic progress is achieved through the alignment of ethical principles with both economic structures and personal choices, as each ethical shift in $Y$ transforms the options available in $D_y$, ultimately shaping individual behavior, economic structures, and societal norms in concert. This layered mapping underscores that sustainable democracy is not only built upon moral principles but also requires the active, ongoing alignment of individual actions with the evolving structure of societal values and economic realities.

In this enriched model, sustainable democracy emerges as an iterative, co-evolving system in which individual choices both shape and are shaped by evolving ethical and economic contexts, fostering a dynamic and resilient trajectory toward a more just, inclusive, and equitable society. The following section applies the vector bundle approach to illustrate this framework in practice.

## Free-grown sugar, ethical consumption, and the limits of materialist economics in 19th-century Britain's unfulfilled path to sustained democracy

In the 1790s, amid a burgeoning moral awakening to the horrors of slavery in Britain, a small group of British entrepreneurs associated with the East India Company introduced a novel market category: "free-grown sugar." This category offered ethically conscious consumers an alternative to slave-produced sugar from the West Indies, acknowledging the moral disquiet around supporting an industry sustained by human bondage (Smith & Johns, 2020; Sussman, 2000). By sourcing sugar from regions where production was free from enslaved labor, these entrepreneurs appealed to a nascent consumer conscience seeking to avoid complicity in the institution of slavery. As Smith and Johns (2020) describe, this initiative marked one of the earliest attempts to create a market shaped by ethical considerations rather than purely economic utility.

Between the 1790s and the 1830s, British retailers provided consumers with a choice between sugar produced through slavery and sugar marketed as "free-grown." Many consumers demonstrated a willingness to pay a premium for the latter, embodying a Kantian deontological commitment that privileged ethical duty over economic convenience. This willingness to act upon moral conviction allowed consumers to express agency through economic decisions, aligning purchasing behavior with ethical values and contributing to a market increasingly reflective of social principles (Mintz, 1986).

The impact of this ethics-driven market extended beyond individual acts of consumption, playing a formative role in shaping public policy and reinforcing the anti-slavery movement. According to Smith and Johns (2020), the success of free-grown sugar as a market force not only served as a precursor to modern fair-trade initiatives but also illustrated how ethical consumption could influence broader economic and social structures. This consumer-driven movement contributed to public support for abolitionist causes, fostering a political climate that would ultimately culminate in the Slavery Abolition Act of 1833 (Trentmann, 2006).

However, after 1840, the ethical market category of free-grown sugar gradually disappeared from Britain, despite continued importation of large quantities of slave-produced sugar. This shift was exacerbated by the 1846 tariff modifications, which controversially lowered barriers for sugar imports from slavery-dependent regions like Cuba and Brazil, where slavery remained legal until the 1880s. These policy changes undercut the demand for ethically sourced sugar by making cheaper, slave-produced sugar more accessible, thereby dismantling the moral economy that free-grown sugar had represented. As Smith and Johns (2020) argue, this prioritization of economic expedience over ethical commitment marked a regression in Britain's moral stance within the global sugar trade, subjugating ethical concerns to material interests and revealing the fragile foothold of ethics-driven markets in the face of economic incentives (Mintz, 1986).

According to Ronald Inglehart's theory of intergenerational value change, Britain in the mid-19th century still operated largely within a materialist framework. Inglehart (1990) argues that societies driven by materialist values prioritize economic stability and physical security over ideals such as social justice and environmental responsibility. This framework posits that shifts toward postmaterialist values, which emphasize ethical and self-expressive concerns, did not begin to take hold in Western societies until the post-World War II era, reaching a turning point from the 1950s onward. Thus, the economic and political structures in mid-19th century Britain were likely more inclined to favor economic expediency over ethical imperatives, contributing to the erosion of the free-grown sugar market in favor of cheaper slave-produced alternatives. Inglehart's theory provides a broader sociocultural explanation for why the ethics-driven consumption of free-grown sugar could not be sustained, as Britain had not yet reached the level of affluence or security that would foster a sustained shift toward postmaterialist values.

The vector bundle approach, particularly through the projection mapping, offers a robust analytical framework to understand when value shifts fail to catalyze sustained democracy. By mapping individual ethical choices and economic structures within a historical context, the projection reveals how value-aligned decisions are filtered through broader material and ethical constraints. In the case of free-grown sugar, the fiber spaces in the bundle suggests that individual consumer choices supporting ethical values were constrained by the economic policies and materialist orientation of 19th-century Britain, which prioritized profit and market efficiency over moral concerns. The projection mapping, in this sense, illustrates that ethical consumerism, while impactful, lacks sufficient structural support in a predominantly materialist society to effect enduring democratic sustainability. For sustainable democratic change to take root, an ethical consumer base must attain a critical mass or "ethical gravity," wherein collective consumer actions meaningfully shape economic structures and policy directions. Reaching this threshold allows ethical values to permeate the democratic fabric, establishing a resilient, value-driven system. The example presented here illustrates the inherent instability of an ethics-driven sugar market within a predominantly materialist economic framework, suggesting that, without a broader societal shift toward post-materialist values, the transition to a sustainable democracy may falter if ethical consumption does not reach a critical mass.

## Conclusion

This paper elucidates the ascendancy of the ethical consumer, a phenomenon deeply entwined with Inglehart's philosophical proposition of intergenerational value shifts, which articulates a societal reorientation toward sustainability and moral responsibility as intrinsic values. The ethical consumer emerges not merely as an economic agent but as a moral arbiter, whose demand for sustainable practices compels institutions and producers to realign their objectives with the imperatives of justice and ecological stewardship. This dynamic reflects the philosophical ethos of sustained democracy, wherein the decentralization of decision-making—exemplified in the general equilibrium model—enshrines the interplay between individual autonomy and the collective good. By invoking principles of transparency, accountability, and mutual responsibility, the rise of the ethical consumer inaugurates a convergence of economic rationality and moral idealism, engendering a resilient social order rooted in ethical and environmental integrity.

This paper argues, echoing Marglin (2023) and Reinke (2021), that mainstream economics should embrace a pluralistic approach. It contrasts the neoclassical model with alternative frameworks that explicitly incorporate ethical values into economic decision-making, emphasizing the growing importance of ethics and moral considerations in economic theory. Specifically, it engages with White's Kantian model of choice (2011), critically examining the ranking axioms that prioritize ethical imperatives within economic contexts.

The analysis identifies that, while White's model introduces a novel perspective, it lacks formal precision, leaving the ranking of ethical preferences somewhat ambiguous. Furthermore, akin to Kant's original framework, White's model does not account for the evolution of moral values over time. To address these limitations, this paper develops foundational criteria for ranking preferences, establishing a more rigorous topological structure for integrating ethics into economic choice.

Additionally, it finds that not all of White's proposed axioms are essential when perfect duties are applied within a discrete topology. By refining these axioms, the paper offers a more robust and adaptable framework for embedding ethical considerations into economic decision-making, providing a path forward for pluralistic and ethically grounded economic models.

With $Y$ as a discrete topology, and $Y \times D$ as the product topology, where $D = X \times E \subseteq \mathbb{R}_+^{n+l}$ carries the standard topology, the projection mapping $\pi: Y \times D \to Y$ is defined on this product topology. This setup provides a framework in which distinct moral and economic phases appear as isolated contexts in $Y$, each associated with a continuous spectrum of choices in $D_{y_i}$, allowing for a nuanced model of ethical and economic decision-making across varied historical and moral landscapes. In this context, the paper has examined Kant's concept of duty in depth, proposing innovative approaches for embedding ethical imperatives within economic decision-making models. Building on this axiomatic foundation, future research may extend these ideas by further analyzing the structure and dynamics of consumer actions within this enriched ethical-economic space.

Though preliminary, the vector bundle approach offers a fresh theoretical framework for examining ethical consumption. Future developments will require formalizing the ethical consumer's optimization problem and deriving the model's equilibrium structure—an endeavor demanding substantial further work including an existence proof of an ethical utility function. Applying degree theory to the projection mapping $\pi: Y \times D \to Y$ introduces a powerful method for examining the multiplicity, stability, orientation, and continuity of equilibrium solutions across various contexts within $Y$. This approach provides a topological lens through which the robustness of moral and economic systems can be analyzed as they evolve through distinct historical phases, navigating both stable configurations and transformative shifts. Future research will build on this foundation, using degree theory to refine the study of ethical and economic transitions within the vector bundle framework and identifying conditions under which core values—such as autonomy and justice—achieve stability and resilience within economic systems.

## Endnotes

[1]    These labels—such as Fair Trade, Organic, and Rainforest Alliance—provide consumers with critical information about the ethical attributes of products, allowing them to make choices that reflect their moral and social values.

[2]    Over time, *homo economicus* became a foundational concept in neoclassical economics, modeling individuals as rational agents who consistently aim to maximize their utility. Despite widespread criticism for its unrealistic assumptions about human behavior, the homo economicus model remains influential in economic theory as a simplification for studying decision-making and market outcomes. Early critiques of *homo economicus* can be traced to Ingram (1888), and Pareto ([1906] 1980), who also helped to popularize the term.

[3]    Debreu's general equilibrium model incorporates uncertain states of nature, making the ranking of preferences state-dependent, as agents adjust preferences based on anticipated conditions (Debreu, 1959). In Debreu's model, the ranking is not over the domain of states themselves but rather over the consumption space associated with each state of nature.

[4]    The Bourbaki group, a collective of French mathematicians, sought to construct a purely abstract mathematical framework, rigorously detached from intuition or empirical grounding (Dieudonné, 1970).

[5]    A preference ordering was initially introduced in Debreu, Gérard (1954), '*Representation of a preference ordering by a numerical function*.' in Robert M. Thrall, Clyde H. Coombs and Robert L. Davis (eds), *Decision Processes*, New York, US: Wiley, pp. 159-165.

[6]    This framework establishes a coherent structure for duty-based preferences without imposing the strict criteria typically required for traditional economic

preferences. Notably, assumptions such as monotonicity, continuity, and convexity remain unaddressed in White (2011).

[7]   Immanuel Kant defines "perfect duties" in his works on moral philosophy, particularly in *Groundwork for the Metaphysics of Morals* (1785) and *The Metaphysics of Morals* ([1797] 2012). In these texts, he differentiates between perfect and imperfect duties based on the strictness and universality of moral obligations.

[8]   In a discrete topological space, defining a constraint as a linear function poses inherent difficulties because the properties that underpin linearity—such as continuity and a well-defined notion of smoothness—are incompatible with the nature of a discrete topology.

[9]   In White's (2011) example, the imperfect duty of charitable giving is portrayed as inherently flexible and adaptable, suggesting how such duties might be conceptualized along a continuous 0–1 scale. Lancaster's *New Consumer Theory* (1966) marks a pivotal philosophical shift from the classical Arrow-Debreu model by redefining the ontology of utility. In the Arrow-Debreu framework, utility is directly tied to the consumption of goods as concrete objects, where the physical item itself is seen as the immediate source of value. Lancaster, however, reconceptualizes this by proposing that utility arises not from the goods themselves, but from the characteristics or attributes they represent. This ontological shift moves the focus of value from the material object to the abstract qualities it embodies, offering a more sophisticated view of consumer preference formation.

[10]  Each slice $\{y_i\} \times \mathbb{R}_+^n$ represents a distinct scenario, where a specific perfect duty or combination of perfect duties is in effect, corresponding to a unique economic-imperfect-duty configuration. The slicing, can be understood as a number of copies of ethical dependent extended Edgeworth boxes. A transition between Edgeworth boxes represents the time parameterized perfect duty bundle shift, representing changes in societal values.

[11]  In contrast to White's axiom, this does not apply a ranking property to perfect duties. Instead, this research considers a time-parameterized evolution of states $y \in Y$, treating these ethical contexts as a sequential progression rather than a hierarchy based on ranking.

[12]  This paper does not extend to a detailed examination of the time-parameterized evolution of perfect duties to capture the historical and institutional context of an economy. This aspect remains a subject for further

research.

## Conflict of interest statement

The author declares that there is no conflict of interest.

## References

Andorfer, V. A. and U. Liebe (2012), 'Research on fair trade consumption—a review', *Journal of Business Ethics*, 106, 415–435.

Andorfer, V. A. and U. Liebe (2015), 'Do information, price, or morals influence ethical consumption? A natural field experiment and customer survey on the purchase of fair trade coffee', *Social Science Research*, 52, 330–350.

Arnold, D. G. and N. E. Bowie (2003), 'Sweatshops and respect for persons', *Business Ethics Quarterly*, 13 (2), 221–242.

Arrow, K. J. and G. Debreu (1954), 'Existence of an equilibrium for a competitive economy', *Econometrica*, 22 (3), 265–290.

Auger, P., Devinney, T. M., Louviere, J. J. and P. F. Burke (2010), 'The importance of social product attributes in consumer purchasing decisions: a multi-country comparative study', *International Business Review*, 19 (2), 140–159.

Basu, Kaushik (2011), *Beyond the Invisible Hand: Groundwork for a New Economics*, Princeton, US: Princeton University Press.

Berlin, Ira (1998), *Many Thousands Gone: The First Two Centuries of Slavery in North America*, Massachusetts, US: Harvard University Press.

Berry, H. and M. McEachern (2005), 'Informing Ethical Consumers', in R. Harrison, T. Newholm, and D. Shaw (eds), *The Ethical Consumer*, London: SAGE, pp. 69–87.

Bowie, Norman E. (1999), *Business Ethics: A Kantian Perspective*, Massachusetts, US: Blackwell.

Bray, J., Johns, N. and D. Kilburn (2011), 'An exploratory study into the factors impeding ethical consumption', *Journal of Business Ethics*, 98 (4), 597–608.

Carrington, M. J., Neville, B. A. and G. J. Whitwell (2010), 'Why ethical consumers don't walk their talk: towards a framework for understanding the gap between the ethical purchase intentions and actual buying behaviour of ethically minded consumers', *Journal of Business Ethics*, 97 (1), 139–158.

Cowe, Roger and Simon Williams (2000), *Who Are Ethical Consumers?*, Manchester, UK: The Co-operative Bank.

Dauvergne, P. (2018), 'Why is the global governance of plastic failing the oceans?', *Global Environmental Change*, 51, 22–31.

Davies, I. A., Lee, Z. and I. Ahonkhai (2021), 'Luxury brands and sustainability: A study on consumers' willingness to pay for sustainability in luxury goods', *Journal of Business Research*, 124, 27–36.

De Pelsmacker, P., Driesen, L. and G. Rayp (2005), 'Do consumers care about ethics? Willingness to pay for fair-trade coffee', *Journal of Consumer Affairs*, 39 (2), 363–385.

Debreu, Gérard (1954), '*Representation of a preference ordering by a numerical function.*' in Robert M. Thrall, Clyde H. Coombs and Robert L. Davis (eds), *Decision Processes*, New York, US: Wiley, pp. 159-165.

Debreu, Gérard (1959), *Theory of Value: An Axiomatic Analysis of Economic Equilibrium*, New Haven, US: Yale University Press.

DeMartino, George F. (2011), *The Economist's Oath: On the Need for and Content of Professional Economic Ethics*, Oxford: Oxford University Press.

Dietz, T., Fitzgerald, A. and R. Shwom (2005), 'Environmental values', *Annual Review of Environment and Resources*, 30, 335–372.

Dieudonné, J. (1970), 'The work of Nicholas Bourbaki', *American Mathematical Monthly*, 77 (2), 134–145.

Dubuisson-Quellier, Sophie (2013), *Ethical Consumption*, translated by Howard Scott, Halifax, Canada: Fernwood Publishing.

Dunlap, R. E. and R. York (2008), 'The globalization of environmental concern and the limits of the postmaterialist values explanation: Evidence from four multinational surveys', *Sociological Quarterly*, 49 (3), 529–563.

Ethical Markets Report 2023 (2024), December 2024. Online at https://www.ethicalconsumer.org/sites/default/files/media-file/2023-12/Ethical-Markets-Report-2023-web-final.pdf (retrieved on 03.11.2024).

Fernández-Ferrín, P., Calvo-Porral, C. and A. Garcia (2023), 'The influence of emotions and social responsibility on willingness to pay for fair trade food products', *Food Quality and Preference*, 106, 104859.

Freestone, O. M. and P. J. McGoldrick (2007), 'Motivations of the ethical consumer', *Journal of Business Ethics*, 79 (4), 445–467.

Friedman, Milton (1953), *Essays in Positive Economics*, London, UK and Chicago, US: University of Chicago Press.

Gandjour, A. (2024), 'Perspectives on interpersonal utility comparisons: an analysis of selected models', Journal of Philosophical Economics, 17, January 2, 2024, https://doi.org/10.46298/jpe.1127.

Genovese, Eugene D. (1974), *Roll, Jordan, Roll: The World the Slaves Made*, New York, US: Pantheon Books.

Gomes, S., Lopes, J. M. and S. Nogueira (2023), 'Willingness to pay more for green products: A critical challenge for Gen Z', *Journal of Cleaner Production*, 390 (136092).

Harper, G. C. and A. Makatouni (2002), 'Consumer perception of organic food production and farm animal welfare', *British Food Journal*, 104, 287-299.

Hausman, D. M. and M. S. McPherson (1993), 'Taking ethics seriously: economics and contemporary moral philosophy', *Journal of Economic Literature*, 31 (2), 671–731.

Heath, J. (2008), 'Business ethics and moral motivation: A criminological perspective', *Journal of Business Ethics*, 83 (4), 595–614.

Hsieh, N. (2004), 'The obligations of transnational corporations: Rawlsian justice and the duty of assistance', *Business Ethics Quarterly*, 14 (4), 643–661.

Hunt, Lynn (2007), *Inventing Human Rights: A History*, New York, US: W.W. Norton & Company.

Inglehart, R. (1971), 'The silent revolution in Europe: intergenerational change in post-industrial societies', *American Political Science Review*, 65 (4), 991–1017.

Inglehart, R. (2008), 'Changing values among western publics from 1970 to 2006', *West European Politics*, 31 (1-2), 130–146.

Inglehart, Ronald (1977), *The Silent Revolution: Changing Values and Political Styles among Western Publics*, Princeton, US: Princeton University Press.

Inglehart, Ronald (1990), *Culture Shift in Advanced Industrial Society*, Princeton, US: Princeton University Press.

Ingram, John K. (1888), *A History of Political Economy*, New York, US: Macmillan and Co.

Kam, C. D. and M. Deichert (2020), 'Boycotting, buycotting and the psychology of political consumerism', *The Journal of Politics*, 82 (1), 72–88.

Kant, Immanuel (1785), *Groundwork of the Metaphysics of Morals*, translated by Herbert J. Paton (1964), New York, US: Harper Torchbooks.

Kant, Immanuel (1797), *The Metaphysics of Morals*, reprinted in M. J. Gregor (ed.) (2012), *Practical Philosophy,* UK: Cambridge University Press.

Lancaster, K. J. (1966), 'A new approach to consumer theory', *Journal of Political Economy*, 74 (2), 132–157.

Lang, T. and Gabriel, Y. (2005), 'A brief history of consumer activism', in Rob Harrison, Terry Newholm, and Deirdre Shaw (eds), *The Ethical Consumer*, London, UK: SAGE, pp. 39–53.

Littrell, Marry A. and Marsha A. Dickson (1999), *Social Responsibility in the Global Market: Fair Trade of Cultural Products*, London, UK: SAGE.

Mai, R. and S. Hoffmann (2015), 'How to combat the unhealthy = tasty intuition: The influencing role of health consciousness', *Food Quality and Preference*, 39, 52–62.

Marglin, S. A. (2008), *The Dismal Science: How Thinking Like an Economist Undermines Community*, Cambridge, US: Harvard University Press.

Marglin, S. A. (2023), 'A plea for pluralism', Journal of Philosophical Economics, 16, 182–195.

Maslow, Abraham H. (1970), *Motivation and Personality*, 2nd ed., New York, US: Harper and Row.

Megicks, P., Memery, J. and J. Williams (2008), 'Influences on ethical and socially responsible shopping: Evidence from the UK grocery sector', *Journal of Marketing Management*, 24 (5/6), 637–659.

Micheletti, M. and D. Stolle (2012), 'Sustainable citizenship and the new politics of consumption', *Annals of the American Academy of Political and Social Science*, 644 (1), 88–120.

Mintz, Sidney W. (1986), *Sweetness and Power: The Place of Sugar in Modern History*, Penguin Books.

Naderi, I. and E. Van Steenburg (2018), "Me first, then the environment: Young millennials as green consumers", *Young Consumers*, 19 (3), 280–295.

New, S. J. (2015), 'Modern slavery and the supply chain: the limits of corporate social responsibility?', *Supply Chain Management*, 20 (6), 697–707.

Newholm, T. and D. Shaw (2007), 'Studying the ethical consumer: A review of research', *Journal of Consumer Behaviour*, 6 (5), 253–270.

Nicholls, A. and Opal, C. (2005), *Fair Trade: Market-Driven Ethical Consumption*, London, UK: SAGE.

Nielsen (2016), 'The Nielsen global responsibility report 2015'. Online at https://www.nielsen.com/us/en/press-room/2015/consumer-goods-brands-that-demonstrate-commitment-to-sustainability-outperform.html (retrieved on 17.10.2024).

Norris, P. and Inglehart, R. (2019), *Cultural Backlash: Trump, Brexit, and Authoritarian Populism*, Cambridge, UK: Cambridge University Press.

Papaoikonomou, E., Ryan, G. and M. Valverde (2011), 'Mapping ethical consumer behavior: integrating the empirical research and identifying future directions', *Ethics & Behavior*, 21 (3), 197–221.

Pareto, Vilfredo (1906), *Manual of Political Economy*, reprinted in A. S. Schwier, and A. Page (eds.) (1980), New York, US: Augustus M. Kelley.

Pepper, M., Jackson, T. and D. Uzzell (2009), 'An examination of the values that motivate socially conscious and frugal consumer behaviours', *International Journal of Consumer Studies*, 33 (2), 126–136.

Reinke, R. (2021), 'A critical note on the scientific conception of economics: claiming for a methodological pluralism', *Journal of Philosophical Economics*, 14(1–2), November 20, 2021, https://doi.org/10.46298/jpe.8664.

Sarkis, Joseph (2023), 'The Circular Economy and Green Supply Chains', in Rico Merkert and Kai Hoberg (eds) *Global Logistics and Supply Chain Strategies for the 2020s.* Cham, Switzerland: Springer.

Sen, A. (1977), 'Rational fools: A critique of the behavioral foundations of economic theory', *Philosophy & Public Affairs*, 6 (4), 317–344.

Sen, Amartya (1985), *Commodities and Capabilities*, India: Oxford University Press.

Smith, A. and J. Johns (2020), 'Historicizing modern slavery: Free-grown sugar as an ethics-driven market category in nineteenth-century Britain', *Journal of Business Ethics*, 166, 271–292.

Stiefenhofer, P. (2019). Conspicuous ethical consumption. *Theoretical Economics Letters*, 9 (1), 1-8.

Stiefenhofer, P. (2021a), 'Towards understanding prices and ethics: Ethical consumers with price-dependent utilities', *Theoretical Economics Letters*, 11 (3), 477–484.

Stiefenhofer, P. (2021b), 'Conspicuous ethics: Existence of price dependent ethical utility functions', *Applied Mathematics*, 12 (4), 252–261.

Stiefenhofer, P. and W. Zhang (2022), 'Conspicuous ethics: A Veblen effect condition for ethical consumption goods', *Applied Economics Letters*, 29 (1), 72–74.

Sussman, Charlotte (2000), *Consuming Anxieties: Consumer Protest, Gender & British Slavery, 1713-1833*, Redwood City, US: Stanford University Press.

Szmigin, I., Carrigan, M. and, M. G. McEachern (2009), 'The conscious consumer: Taking a flexible approach to ethical behaviour', *International Journal of Consumer Studies*, 33 (2) 224–231.

Tan, C.N.L., Ojo, A.O. and Thurasamy, R. (2019), 'Determinants of green product buying decision among young consumers in Malaysia', *Young Consumers*, 20 (2), 121-137.

Trentmann, Frank (ed) (2006, *The Making of The Consumer: Knowledge, Power and Identity in The Modern World,* Cultures of Consumption Series, Oxford, UK: Berg Publishers.

Trudel, R. and J. Cotte (2009), 'Does it pay to be good?', *MIT Sloan Management Review*, 50 (2), 61–68.

Utt, Ronald D. (1991), 'Privatization in the United States', in Attiat F. Ott and Keith Hartley (eds), *Privatization and Economic Efficiency*, Aldershot, UK and Brookfield, US: Edward Elgar, pp. 73–86.

Veblen, Thorstein (1899), *The Theory of the Leisure Class: An Economic Study of Institutions*, New York, US: Macmillan.

Vredenburg, J., Kapitan, S., Spry, A. and J. A. Kemper (2020), 'Brands taking a stand: Brand activism as a key business strategy', *Journal of Business Research*, 116, 163–170.

Wells, V., Ellis, N., Slack, R. and M. Moufahim (2019), '"It's us, you know, there's a feeling of community": Exploring notions of community in a consumer co-operative', *Journal of Business Ethics*, 158, 617–635.

Wettstein, F., Giuliani, E., Santangelo, G. D. and G. K. Stahl (2019), 'International business and human rights: A research agenda', *Journal of World Business*, 54 (1), 54-65.

White, K., Habib, R. and D. J. Hardisty (2019), 'How to shift consumer behaviors to be more sustainable: A literature review and guiding framework', *Journal of Marketing,* 83 (3), 22-49.

White, Mark D. (2011), *Kantian Ethics and Economics: Autonomy, Dignity, and Character*, Stanford, US: Stanford University Press.

Pascal Stiefenhofer is a senior lecturer in the Department of Economics, Newcastle University Business School (UK) (pascal.stiefenhofer@newcastle.ac.uk).